\documentclass[12pt, fleqn]{article}

\begin{document}

\title{Questions on the concept of time}

\author{Luigi Foschini\\
{\small CNR--Institute FISBAT, Via Gobetti 101, I--40129, Bologna (Italy)}\\
{\small Email: \texttt{L.Foschini@fisbat.bo.cnr.it}}}

\maketitle

\begin{abstract}
	Some notes and questions about the concept of time are exposed. 
	Particular reference is given to the problem in quantum mechanics, in 
	connection with the indeterminacy principle.
\end{abstract}

03.65.Bz Foundations, theory of measurement, miscellaneous theories.

\vskip 24pt

``How much time'', ``It took quite some time'', ``There is plenty of 
time''.  These are some examples only of our common way to think about 
time: an interval between two instants.  In physics as well time is 
seen as an interval.  Asher Peres stressed that the measurement of 
time is the observation of a dynamical variable, which law of motion 
is known and it is uniform and constant in time \cite{PERES}.  There is a 
kind of self--reference in this definition and nothing is said about 
time.  On the other hand, time is considered simply as a parameter 
and, according to this, the above definition is completely 
satisfactory.  Moreover, time is sometimes neglected (e.g.  steady 
state phenomena), which is useful to understand some physical 
concepts.

However, when we deal with quantum mechanics the problem of the time 
explodes in all its complexity.  In classical mechanics (hamiltonian 
formulation) the dynamical state of a physical system is described by 
a point in a phase space, that is we have to know position $q$ and 
momentum $p$ at a given time $t$.  Even though it can appear a 
sophism, it is not possible, strictly speaking, to know simultaneously 
$q$ and $p$ of any object.  However, in classical physics, we may 
neglect variations during the lapse of time between the measurement of 
$q$ and $p$, because the quantum of action is so small when compared 
to macroscopic actions.  It is very interesting to note how Sommerfeld 
stressed this question when he wrote about the Hamilton's principle of 
least action: the trajectory points $q$ and $q+\delta q$ are 
considered \emph{at the same time instant} \cite{SOMMERFELD}.

In quantum mechanics this approximation is not valid, because actions 
are comparable with the quantum of action.  The hamiltonian formalism 
is not anymore a useful language to investigate nature and, as known, 
it was necessary to settle quantum mechanics.  

The impossibility to neglect time in quantum mechanics is well described 
by the Heisenberg's principle of indeterminacy. Nevertheless, the role of 
the time in quantum indeterminacy is often neglected. In the history 
of physics, we can often find authors which claimed to 
have found a way to avoid the obstacle of indeterminacy. However they all 
missed the target, that is the question of the time.  Heisenberg clearly 
stated that indeterminacy relationships do not allow a simultaneous 
measurement of $q$ and $p$, while do not prevent from measuring $q$ and 
$p$ taken in isolation \cite{HEISENBERG}. It is possible to measure, 
with great precision, complementary observables in two different time 
instants: this is not forbidden by Heisenberg's principle.  Later 
on, it is also possible to reconstruct one of the observables at the reference 
time of the other observable, but this is questionable.  In the interval between 
two measurements observables change because time flows.  What 
happened during this interval?  We can reconstruct observables by making 
hypoteses, but we have to remember that these are hypoteses and not 
measurements.

We have to take into account the so--called ``energy--time uncertainty 
relationship''.  As known, time is a $c$--number and therefore it have 
to commute with each operator.  Nevertheless, the relationship exists, 
but it is worth to note its dynamical nature, whereas indeterminacy is 
kinematic \cite{AHARONOV}.  That is, it follows from the evolution of 
the system during the measurement. Bohr had already stated this and he had 
often pointed out the time issue \cite{BOHR}, along with 
Landau and Peierls \cite{LANDAU}.  We refer to \cite{LANDAU} in 
which the question is stated in a better way.  The relationship:

\begin{equation}
	\Delta E \Delta t > \hbar
	\label{e:energy}
\end{equation}

\noindent means that we have to consider the system evolution  
during the measurement, that is the difference between the 
measurement result and the state after the measurement.  The 
energy difference between the two states cannot be less than 
$\hbar/\Delta t$.  The energy--time relationship has important 
consequences particularly as regards the momentum measurement and, 
therefore, on double--slit experiment \cite{LANDAU}.

Eq.~(\ref{e:energy}) suggests that, given a certain 
energy, it is possible to construct a state with a huge $\Delta E$ in order 
to obtain a very small $\Delta t$. However, in a recent paper, Margolus and Levitin
\cite{MARGOLUS} give a strict bound that depends on the difference between the average
energy of the system and its ground state energy. Is it a step toward a quantization
of the time?

In addition, if we consider the equation of motion (written with 
Dirac's notation \cite{DIRAC}):

\begin{equation}
	i\hbar \frac{d|Pt>}{dt}=H(t)|Pt>
	\label{e:moto}
\end{equation}

\noindent we can see that $H(t)$ is $i\hbar$ times an operator of 
time--translation.  If the system is closed we can consider $H$ 
constant and equal to the total energy of the system; but if not, if 
energy depends on time, this means that the system is under the action 
of external forces (e.g.  measurement).  The measurement introduce an 
energy exchange that does not follow causality.

Moreover, it is worth to note that a closed system is an 
abstraction.  A real closed system is not observable, without 
introducing energy exchange which would change $H$: therefore that would not 
be a closed system. We can say, by means of Rovelli's words that there is no way to
get information about a system without physically interacting with it for a certain
time \cite{ROVELLI}.

Would you consider it a sophism?  Of course not. We should always bear in mind that 
quantum physics is only an interpreted language we use to speak about 
Nature, though it does not describe Nature itself (on logic--linguistic structure 
of quantum physics see, for example, \cite{FOSCHINI}).  In classical 
physics we made many approximations, which are no longer valid in 
quantum physics. In particular, we can no more neglect time. 
As Heraclitus stated, you cannot plunge your hands twice in the same stream.

\end{document}